\documentclass[twocolumn,showpacs,prl]{revtex4}

\usepackage{amsmath2000,amssymb}
\usepackage{graphicx,bm}
\usepackage{epsfig}
\newcommand{\Tr}{{\rm Tr}}
\newcommand{\tr}{{\rm tr}}

\begin{document}

\title{Effects of spin and exchange interaction on the Coulomb-blockade
peak statistics in quantum dots}

\author{Y.\ Alhassid and T.\ Rupp}
\affiliation{Center for Theoretical Physics, Sloane Physics
  Laboratory, Yale University,  New Haven, Connecticut 06520, USA}

\begin{abstract}
  We derive a closed expression for the linear conductance through a
  quantum dot in the Coulomb-blockade regime in the presence of a
  constant exchange interaction. With this expression we calculate
  the temperature dependence of the conductance peak-height and
  peak-spacing statistics. Using a realistic value of the exchange
  interaction, we find significantly better agreement with
  experimental data as compared with the statistics obtained in the
  absence of an exchange interaction.
\end{abstract}
\pacs{73.23.Hk, 05.45.Mt, 73.40.Gk, 73.63.Kv}

\maketitle

The conductance through a quantum dot that is weakly coupled to leads
displays sharp peaks as an applied gate voltage is varied.  Each
conductance peak describes the addition of one more electron into the
dot, and between peaks the conductance is ``blocked'' by the Coulomb
interaction. The statistics of both peak heights and peak spacings in
dots for which the single-electron dynamics is chaotic have been
intensively studied in recent years~\cite{alhassid00}. Some of the
experimental observations, e.g.\ the peak-height distributions at low
temperature~\cite{jalabert92,chang96,folk96}, could be explained at
least qualitatively by the constant-interaction (CI) model, in which
the interaction is represented in the simple form of an electrostatic
charging energy. Other measured observables, such as the peak-spacing
distribution~\cite{patel98a}, have indicated that spin and residual
interactions beyond the charging energy should be taken into account.
A consistent theoretical approach that provides quantitative agreement
with both the measured peak-height and peak-spacing statistics is
still lacking.

Recently, a universal Hamiltonian was
derived~\cite{kurland00,aleiner02} for a dot with a large Thouless
conductance $g_{\rm T} \sim \sqrt{N}$ ($N$ is the number of
electrons). An important contribution to the interaction part of this
Hamiltonian is a constant exchange interaction in addition to the
usual charging-energy term. The remaining interaction terms are
suppressed at large $g_{\rm T}$.  Here we study the effect of the
exchange interaction on the finite-temperature statistics of both peak
heights and spacings.  To this end, we derive a closed expression for
the conductance in the presence of a constant exchange interaction (in
the sequential-tunneling limit).  This formula expresses the
conductance in terms of quantities that characterize spinless
non-interacting electrons.  We then calculate the finite-temperature
peak-height and peak-spacing statistics and find them both to be
sensitive to the exchange interaction.  Using an RPA estimate of the
exchange interaction for the samples studied experimentally in
Refs.~\cite{patel98a,patel98b}, we obtain very good agreement with the
observed temperature dependence of the standard deviation of the peak
spacing. We also explain most of the known discrepancies between the
experimental peak-height statistics~\cite{patel98b} and the
predictions of the CI model for $kT \lesssim 0.6\ \Delta$ ($\Delta$ is
the mean spacing between spin-degenerate levels).

The universal Hamiltonian of a quantum dot in the limit $g_{\rm T} \to
\infty$ is given by~\cite{kurland00,aleiner02}
\begin{equation}\label{universal}
{\hat H} = \sum_{\lambda \sigma}\epsilon_\lambda a^\dagger_{\lambda
\sigma}a_{\lambda \sigma} + \frac{e^2}{2C} {\hat n}^2 - J_s {\bf \hat
S}^2\;,
\end{equation}
where $\epsilon_\lambda$ are spin-degenerate single-particle levels
($\sigma = \pm 1$ labels the spin). The second term in
Eq.~(\ref{universal}), where $C$ is the dot's capacitance and $\hat n$
is the total--particle-number operator, accounts for the electrostatic
energy of the dot.
The third term, in which ${\bf \hat S}$ is the total-spin operator,
describes a constant exchange interaction with strength
$J_s$. The
occupation-number operator $\hat n_\lambda=\hat n_{\lambda+} + \hat
n_{\lambda -}$ of any single-particle orbital $\lambda$ commutes with
the total spin, $[\hat n_\lambda, {\bf \hat S}]=0$, and the
Hamiltonian $\hat H$ is invariant under spin rotations.  Thus the
eigenstates of $\hat H$ are characterized by their particle number
$N$, the configuration of orbital occupation numbers ${\bf n} =
\{n_\lambda\}$ ($n_\lambda=0,1$ or $2$), the total spin $S$, and the
spin projection $S_z=M$. We label the eigenstates as $|N {\bf n} \gamma S
M\rangle$ where the quantum number $\gamma$ distinguishes between
states with the same total spin $S$ and particle configuration ${\bf
  n}$. The eigenenergies are given by $\varepsilon_{{\bf
    n} S}^{(N)}= \sum_\lambda \epsilon_\lambda n_\lambda+e^2 N^2/2C -
J_s S(S+1)$.

In the limit of sequential tunneling (when a typical tunneling width
is small compared with  $kT$ and $\Delta$), the conductance can be
calculated using a rate-equations approach.  In
Ref.~\cite{alhassid02a}, we developed such an approach in the
presence of interactions and spin. In particular, an explicit solution
exists when the orbital occupation numbers $n_\lambda$ are good quantum numbers.
Expressing the conductance $G$ in a rescaled form $G= (e^2 \bar
\Gamma/8\hbar kT) g$ (where $\bar \Gamma$ is an average width of a level), we have,
in the vicinity of the $N\!+\!1$-st Coulomb-blockade peak
\begin{equation}\label{closed-conductance}
\!\! g=4\!\!\!\sum_{\lambda {\bf n} \gamma S \atop {\bf n'} \gamma' S'}\!\!\!\!
{\tilde P}_{{\bf n} S}^{(N)} f(\varepsilon_{S'S}^\lambda) |([N\!+\!1]
{\bf n'} \gamma' S' |\!| a_\lambda^\dag |\!| N {\bf n} \gamma S)|^2
g_\lambda .
\end{equation}
Here
$g_\lambda\! =\! 2 {\bar\Gamma}^{-1} \Gamma_\lambda^{\rm l} \Gamma_\lambda^{\rm r}/
    (\Gamma_\lambda^{\rm l} + \Gamma_\lambda^{\rm r})$
are the single-particle level conductances, where $\Gamma^{\rm
  l,r}_\lambda$ are the partial widths of an electron in orbital
$\lambda$ to decay to the left or right lead. The equilibrium
probability of the dot to be in the state $|N {\bf n}\gamma S
M\rangle$ is ${\tilde P}_{{\bf n} S}^{(N)} =e^{-\beta(\epsilon_{n
    S}^{(N)} - {\tilde \epsilon}_{\rm F}N)}/Z$, where the partition
function $Z$ is a Boltzmann-weighted sum over all possible $N$- and
$(N+1)$-body states (no other particle numbers contribute because of
the charging energy), and ${\tilde \epsilon}_{\rm F} = e \zeta V_{\rm
  g} + \epsilon_{\rm F}$ is an effective Fermi energy ($\epsilon_{\rm
  F}$ is the Fermi energy in the leads, $V_{\rm g}$ is the gate
voltage and $\zeta=C_{\rm g}/C$ with $C_{\rm g}$ the dot-gate
capacitance). The Fermi-Dirac function $f(x)=(1+e^{\beta x})^{-1}$ is
evaluated at an electron energy (relative to the Fermi energy)
$\varepsilon_{S'S}^\lambda = \varepsilon_{{\bf n'}S'}^{(N+1)} -
\varepsilon_{{\bf n}S}^{(N)} - {\tilde \epsilon}_{\rm F}$ that
conserves energy at the transition between states $|N{\bf n}\gamma S
M\rangle$ and $|(N+1){\bf n'}\gamma'S'M'\rangle$. The corresponding
reduced matrix element $([N\!+\!1] {\bf n'} \gamma' S' |\!|
a_\lambda^\dag |\!| N {\bf n} \gamma S)$ enforces the selection rule
$S'=|S \pm 1/2|$.

Eq.~(\ref{closed-conductance}) can be rewritten in the form
\begin{equation}\label{conductance}
g = \sum_\lambda (w_\lambda^{(0)} + w_\lambda^{(1)}) g_\lambda \;,
\end{equation}
where the contributions with $n_\lambda = 0$ and $n_\lambda=1$ are
collected in $w_\lambda^{(0)}$ and $w_\lambda^{(1)}$, respectively.
For the cases with $n_\lambda=0$, the final $(N\!+\!1)$-particle state
is given by $|(N\!\!+\!\!1){\bf n'} \gamma' S' M'\rangle\!=\!\sum_{M
  \sigma} (SM{\textstyle \frac{1}{2}} \sigma|S'M')
a_{\lambda\sigma}^\dag|N{\bf n}\gamma S M \rangle$, where
$(SM{\textstyle \frac{1}{2}} \sigma|S'M')$ is a Clebsch-Gordon
coefficient.  When $n_\lambda=1$ (and hence $n'_\lambda=2$), the
$N$-particle state can be similarly related to the $(N+1)$-particle
state by changing to a hole representation. This leads to the
following reduced matrix elements,
\begin{equation}
\!({\bf n'}\gamma'S'|\!|a_\lambda^\dag|\!|{\bf n}\gamma S)\! =
\!(-)^{S-S'-\frac{1}{2}}\! \left\{
\begin{array}{ll}
\!\!\sqrt{2S'\!+\!1} & {\rm if}\ n_\lambda\!=\!0,\\
\!\!\sqrt{2S\!+\!1}  & {\rm if}\ n'_\lambda\!=\!2.
\end{array}
\right.
\end{equation}
Using the relation ${\tilde P}_{{\bf n} S}^{(N)}
f(\epsilon_{S'S}^\lambda) = {\tilde P}_{{\bf n'} S'}^{(N+1)}
[1\!-\!f(\epsilon_{S'S}^\lambda)]$, we get
\begin{subequations}
\label{w_spinproj}
\begin{equation}
w_\lambda^{(0)} = 4\sum_S b_{\lambda,N,S} P_{N,S}\!\!\!
\sum_{S'=S\pm 1/2}\!\!\! (2S'\!+\!1) f(\epsilon_{S'S}^\lambda)\,,
\label{w0_spinproj}
\end{equation}
\begin{equation}
w_\lambda^{(1)} = 4\sum_{S'} c_{\lambda,N+1,S'} P_{N+1,S'}\!\!\!
\sum_{S=S'\pm 1/2}\!\!\! (2S\!+\!1) [1\!-\!f(\epsilon_{S'S}^\lambda)]\,,
\label{w1_spinproj}
\end{equation}
\end{subequations}
where the quantities $b_{\lambda,N,S}= \frac{1}{2} \langle({\hat
  n}_\lambda\!-\!1) ({\hat n}_\lambda\!-\!2)\rangle_{N,S}$ and
$c_{\lambda,N,S}= \frac{1}{2} \langle {\hat n}_\lambda ({\hat
  n}_\lambda\!-\!1) \rangle_{N,S}$ ensure that the sum is only over
contributions with $n_\lambda=0$ or 1, respectively. They are defined
in terms of thermal expectation values at constant particle number $N$
and spin $S$, i.e.\ $\langle {\hat X} \rangle_{N,S} = \Tr_{N,S} [{\hat
  X} e^{-\beta {\hat H}}]/\Tr_{N,S} [e^{-\beta {\hat H}}]$. The
quantity $P_{N,S}$ is the probability to find the dot with $N$
electrons and spin $S$,
\begin{equation}\label{spin-prob}
P_{N,S} = e^{-\beta [F_{N,S} + U_{N,S}]}/ Z \;,
\end{equation}
where $F_{N,S} = -\beta^{-1} \ln {\rm Tr}_{N,S} e^{-\beta
  \sum_{\lambda\sigma} \epsilon_\lambda a^\dag_{\lambda\sigma}
  a_{\lambda\sigma}}$ is the free energy of $N$ non-interacting
electrons with total spin $S$ and $U_{N,S} = e^2 N^2/2C - J_s S(S+1) -
\tilde \epsilon_{\rm F} N$.

The spin-projected trace of a scalar observable can be calculated from
traces at fixed spin projection $M$ using $\Tr_{N,S} \hat X =
\Tr_{N,M=S} \hat X - Tr_{N,M=S+1} \hat X$.  For spin-$1/2$ particles,
the projection on fixed particle number $N$ and spin projection $M$ is
equivalent to projecting on a fixed number of spin-up and spin-down
particles $n_\pm = N/2 \pm M$.
Therefore,
\begin{equation}
\Tr_{N,S} {\hat X} = \tr_{\frac{N}{2}+S,\frac{N}{2}-S} {\hat X} -
\tr_{\frac{N}{2}+S+1,\frac{N}{2}-(S+1)} {\hat X}\,,
\label{eq_spintrace}
\end{equation}
where the traces ``$\tr$'' on the r.h.s.~are evaluated at fixed $n_+$
and $n_-$.  Using ${\hat X}= e^{-\beta \sum_{\lambda\sigma}
  \epsilon_\lambda a^\dag_{\lambda\sigma} a_{\lambda\sigma}}$ in
Eq.~(\ref{eq_spintrace}) we find that the free energy in
Eq.~(\ref{spin-prob}) is given by
\begin{equation}
e^{\!\!-\beta F_{N,S}}\!=\!e^{\!\!-\beta({\tilde F}_{\frac{N}{2}+S}\!+{\tilde
F}_{\frac{N}{2}-S})} \!\!\!-\! e^{\!\!-\beta({\tilde F}_{\frac{N}{2}+S+1}\!
+{\tilde F}_{\frac{N}{2}-(S+1)})}\!\!.
\label{sp_freeenergy}
\end{equation}
The free energy ${\tilde F}_q$ in Eq.~(\ref{sp_freeenergy}) is defined
for $q$ spinless particles $e^{-\beta {\tilde F}_q} = {\rm \tilde
  tr}_q e^{-\beta \sum_\lambda \epsilon_\lambda c_\lambda^\dag
  c_\lambda}$ where $c_\lambda^\dag$ and $c_\lambda$ create and
annihilate spinless particles in non-degenerate levels with energies
$\epsilon_\lambda$.  The quantity $c_{\lambda,N,S}$ from
Eq.~(\ref{w1_spinproj}) can now be expressed as
\begin{widetext}
\begin{equation}\label{c-function}
c_{\lambda,N,S} = \frac{\langle {\tilde n}_\lambda
\rangle_{\frac{N}{2}+S} \langle {\tilde n}_\lambda \rangle_{\frac{N}{2}-S}
e^{-\beta (\tilde F_{\frac{N}{2}+S} + \tilde F_{\frac{N}{2}-S})}
- \langle {\tilde n}_\lambda \rangle_{\frac{N}{2}+S+1}
\langle {\tilde n}_\lambda \rangle_{\frac{N}{2}-(S+1)}
e^{-\beta (\tilde F_{\frac{N}{2}+S+1} + \tilde F_{\frac{N}{2}-(S+1)})}}
{e^{-\beta(\tilde F_{\frac{N}{2}+S} +\tilde  F_{\frac{N}{2}-S})} -
e^{-\beta(\tilde F_{\frac{N}{2}+S+1} + \tilde F_{\frac{N}{2}-(S+1)})}}
\end{equation}
\end{widetext}
where $\tilde n_\lambda$ is the particle-number operator of a
non-degenerate orbital $\lambda$. The function $b_{\lambda,N,S}$ from
Eq.~(\ref{w0_spinproj}) is expressed by replacing $\tilde n_\lambda$
by $(1-\tilde n_\lambda)$ in Eq.~(\ref{c-function}). The complete
expression for the conductance is then obtained from
Eqs.~(\ref{conductance}), (\ref{w_spinproj}), (\ref{spin-prob}),
(\ref{sp_freeenergy}), (\ref{c-function}) and the relation indicated
in the previous sentence. Thus the dot's conductance in model
(\ref{universal}) is determined in terms of the free energy $\tilde
F_q$ and single-particle occupation numbers $\langle \tilde
n_\lambda\rangle_q$ of $q$ non-interacting spinless fermions.  Both
${\tilde F}_q$ and $\langle\tilde n_\lambda\rangle_q$ are familiar
from earlier works in the framework of the CI model, and can be
expressed in closed form using particle-number projection [see
Eqs.~(140) in Ref.~\cite{alhassid00}].

In chaotic dots, the single-particle Hamiltonian in (\ref{universal})
is described by random-matrix theory. We have studied the statistics
of peak heights and spacings for both the orthogonal and unitary
symmetries. The dimension of the configuration sum in
Eq.~(\ref{closed-conductance}) increases combinatorially with the
number of single-particle orbitals and a direct use of
(\ref{closed-conductance}) becomes impractical at higher temperatures.
In contrast, the closed expression we derived greatly facilitates the
calculation of the conductance for a rather large model system of 50
single-particle orbitals $\lambda$. We checked that our results are
not affected by the finite size of the system up to temperatures of
$kT \sim 3\ \Delta$.

\begin{figure}
\epsfxsize=0.75\columnwidth
\epsfbox{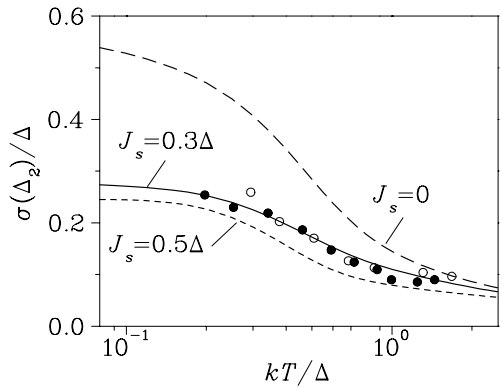}
\caption{The width $\sigma(\Delta_2)$ of the peak-spacing distribution
  for three different values of the exchange-interaction strength
  $J_s$. The symbols are the experimental data of
  Ref.~\cite{patel98a}.}
\label{fig1}
\end{figure}

Theoretical calculations of the width $\sigma(\Delta_2)$ of the
peak-spacing distribution, based on a spinless CI
model~\cite{alhassid99}, describe qualitatively the
observed decrease of this quantity with increasing
temperature~\cite{patel98a}. However, a proper modeling of the
peak-spacing distribution itself requires the inclusion of spin. When
spin is included and in the absence of an exchange interaction, the
calculated values of $\sigma(\Delta_2)$ (long-dashed line in
Fig.~\ref{fig1}) show a large discrepancy with the experimental values
(symbols). Fig.~\ref{fig1} also shows $\sigma(\Delta_2)$ for non-zero
values of $J_s$. For a gas constant of $r_s\sim
1.2$ (that corresponds to the samples used in the experiments), the
RPA estimate is $J_s \approx 0.3\ \Delta$~\cite{oreg02}, and we find
for this value a very good agreement with the measurements.  The results
for $J_s=0.5\ \Delta$ underestimate
the experimental widths. We remark that at temperatures $kT \lesssim
0.4\ \Delta$, the model (\ref{universal}) does not describe well
the shape of the peak-spacing distribution, and it is necessary to
include the fluctuating part of the universal Hamiltonian to explain
the absence of bimodality~\cite{usaj01,alhassid02}. At higher
temperatures, the bimodality is absent already in model
(\ref{universal}) and the residual interaction has a negligible
effect on the width.

\begin{figure}
  \epsfxsize=0.9\columnwidth \epsfbox{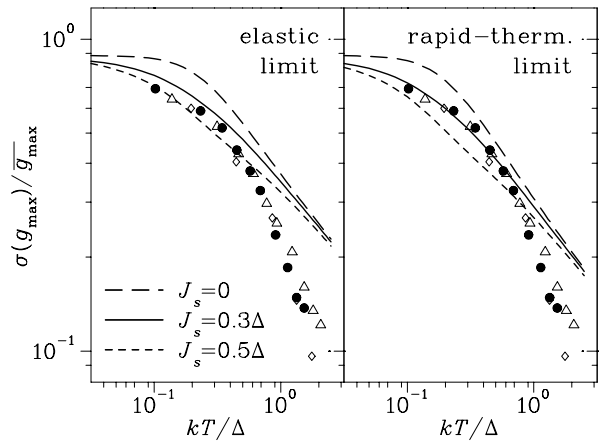}
\caption{The ratio $\sigma(g_{\rm max})/\bar{g}_{\rm max}$
  between the standard deviation and the average value of the peak
  height versus temperature $kT$. The left (right) panel shows data in
  the case of the elastic (rapid-thermalization) limit for three
  different strengths of the exchange interaction $J_s=0$
  (long-dashed), $0.3\ \Delta$ (solid), and $0.5\ \Delta$
  (short-dashed). The symbols are the experimental data of
  Ref.~\cite{patel98b}.}
\label{fig2}
\end{figure}

Another measured quantity is the ratio
between the standard deviation $\sigma(g_{\rm max})$ and the average
$\bar{g}_{\rm max}$ of the peak heights $g_{\rm
  max}$~\cite{patel98b}. The experimental data for this ratio (symbols
in Fig.~\ref{fig2}) are seen to be suppressed in comparison with the
results of model (\ref{universal}) without an exchange term
(long-dashed line in the left panel of Fig.~\ref{fig2}). Spin-orbit
interaction was proposed as a mechanism for this suppression at low
temperatures~\cite{held02}. It was necessary to assume a spin-orbit
coupling that is sufficiently strong to completely decorrelate the spin-up and
spin-down levels. However, spin-orbit effects are likely to be
suppressed in the small dots used in the experiment.  To determine
whether an exchange interaction can explain the observed suppression
of $\sigma(g_{\rm max})/\bar{g}_{\rm max}$, we calculated this
ratio versus temperature $kT$ for different strengths of the exchange
interaction (see Fig.~\ref{fig2}). In the elastic limit (left panel),
a realistic exchange interaction of $J_s=0.3\ \Delta$ leads to closer
agreement with the data. The remaining small discrepancy at
temperatures $kT \lesssim 0.6\ \Delta$ can probably be accounted for
by adding a realistic weak spin-orbit interaction. It still remains
to explain the discrepancy at higher temperatures, where inelastic
 scattering may play a role. The calculation of
Ref.~\onlinecite{rupp02} showed that the suppression of $\sigma(g_{\rm
  max})/\bar{g}_{\rm max}$ due to inelastic scattering is small
for $J_s=0$. In the right panel of Fig.~\ref{fig2} we show results
for the rapid-thermalization limit of strong inelastic scattering in
the presence of an exchange interaction.  While the agreement (for
$J_s=0.3\ \Delta$) is now better at low temperatures, we do not expect
inelastic scattering to be important at these temperatures.  At higher
temperatures, the rapid-thermalization limit does not describe the
data, and it would be interesting to determine the effect of an additional
 weak spin-orbit term.
\begin{figure}
\psfig{figure=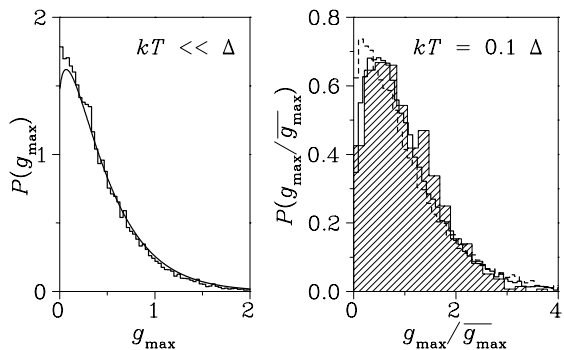,width=0.52\columnwidth,angle=90}
\caption{Peak-height distributions. Left panel: The analytically known
  distribution $P(g_{\rm max})$ at $kT \ll \Delta$ and $J_s=0$ (smooth
  curve) is compared with the corresponding distribution at
  $kT=0.01~\Delta$ and $J_s=0.5~\Delta$ (histogram).  Right panel:
  Experimental data from Ref.~\cite{patel98b} (gray-shaded histogram)
  at $kT=0.1~\Delta$ are compared with the calculated distributions at
  $J_s=0$ (dashed histogram) and $J_s=0.3~\Delta$ (solid histogram).}
\label{fig3}
\end{figure}

For $kT \ll \Delta$ and $J_s=0$, the peak-height distribution
$P(g_{\rm max})$ can be calculated analytically~\cite{jalabert92} and
is shown for the unitary symmetry as a solid line in the left panel of
Fig.~\ref{fig3} (compared to the case of spinless electrons, the peak
heights are rescaled~\cite{alhassid00} by $8(\sqrt{2}-1)^2 \sim
1.37$).  Also shown (histogram) is the peak-height distribution
calculated at $kT=0.01\ \Delta$ and $J_s=0.5\ \Delta$. No
significant effect due to exchange is
observed except for a small enhancement of the probability at small
peak heights.

At finite temperature, the exchange interaction has a stronger effect
on the peak-height distribution. The right panel of Fig.~\ref{fig3}
compares the histogram (gray shaded) of the experimental data for
$P(g_{\rm max}/\bar{g}_{\rm max})$ at $kT=0.1\ \Delta$ with the
calculated histograms for the cases of no exchange ($J_s=0$) and
$J_s=0.3\ \Delta$. This latter realistic value of the exchange
interaction explains the observed suppression of the probability at
small peak heights.

\begin{figure}
\psfig{figure=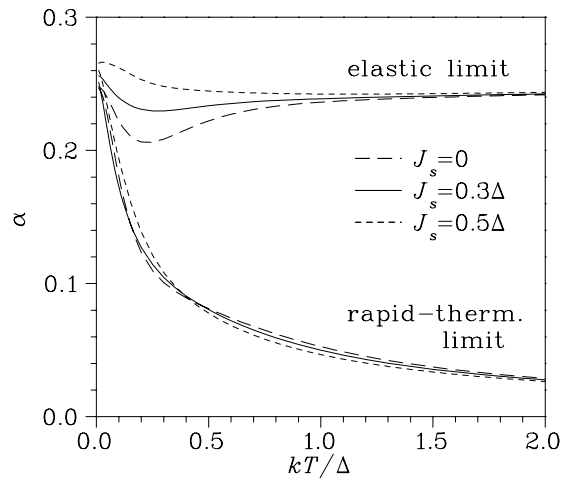,width=0.72\columnwidth,angle=90}
\caption{The weak-localization parameter $\alpha$ versus temperature
  $kT$ in the elastic and rapid-thermalization limit for three
  different values of the exchange-interaction strength $J_s$.}
\label{fig4}
\end{figure}

The weak-localization effect in the average peak height attracted
recent attention both in experiment and
theory~\cite{folk01,beenakker01,eisenberg02,rupp02}. Its suppression at
higher temperatures was suggested as a signature of inelastic
scattering in the dot. The effect is quantified by the parameter
\mbox{$\alpha\! =\! 1\! -\! (\bar{g}_{\rm max}^{\rm GOE}/
  \bar{g}_{\rm max}^{\rm GUE})$.}
In the rapid-thermalization
limit, $\alpha$ decreases rapidly with increasing temperature from its
value of $0.25$ at $kT \ll \Delta$~\cite{beenakker01}. In contrast, if inelastic
scattering is negligible, $\alpha$ was expected to be temperature
independent.  However, calculations for $J_s=0$ showed a slight
suppression of the elastic $\alpha$ around $kT \sim
0.25~\Delta$~\cite{rupp02,eisenberg02}. This was understood by the fact
that close lying levels and hence higher conductances are more likely
for the orthogonal symmetry. The effect of the exchange interaction on
$\alpha$ is shown in Fig.~\ref{fig4}. We find that the dip in $\alpha$
around $kT \sim 0.25~\Delta$ is flattened out and in the
small-temperature limit $\alpha$ becomes larger than $0.25$.  While
$\alpha$ is seen to be sensitive to the exchange interaction at low
temperatures, the experimental uncertainties of Ref.~\cite{folk01} are
too large to observe this effect. In the rapid-thermalization limit,
$\alpha$ is insensitive to $J_s$.

In conclusion, we have derived a closed expression for the conductance in
the presence of spin and exchange interaction. Using this formula we
studied the dependence of the peak height and spacing statistics on
the exchange interaction and found a significantly better quantitative
agreement with experiment than in the absence of an exchange interaction.

This work was supported in part by the U.S. DOE grant No.\
DE-FG-0291-ER-40608. We thank C.~M.\ Marcus for helpful discussions.

\vfill

\end{document}